\documentclass[twocolumn,showpacs,preprintnumbers,amsmath,amssymb,nofootinbib,superscriptaddress]{revtex4-1}

\usepackage[utf8]{inputenc}
\usepackage{graphicx}
\usepackage{amssymb}
\usepackage{textcomp}
\usepackage{amsmath}
\usepackage{tabularx}
\usepackage{bm}
\usepackage{times}
\usepackage{color}
\usepackage{slashed}
\usepackage{multirow}
\usepackage{verbatim}
\usepackage{cancel}
\usepackage{subfigure}

\usepackage[colorlinks=true, pdfstartview=FitV, linkcolor=blue, citecolor=blue, urlcolor=blue]{hyperref}
\allowdisplaybreaks[4]

\def\lsim{\mathrel{\raise.3ex\hbox{$<$\kern-.75em\lower1ex\hbox{$\sim$}}}}
\def\gsim{\mathrel{\raise.3ex\hbox{$>$\kern-.75em\lower1ex\hbox{$\sim$}}}}

\definecolor{orange}{rgb}{1,0.5,0}

\begin{document}

\title{Electron-target experiment constraints on light dark matter produced in primordial black hole evaporation}

\author{Tong Li}
\email{litong@nankai.edu.cn}
\affiliation{
School of Physics, Nankai University, Tianjin 300071, China
}
\author{Jiajun Liao}
\email{liaojiajun@mail.sysu.edu.cn}
\affiliation{
School of Physics, Sun Yat-Sen University, Guangzhou 510275, China
}

\begin{abstract}
Light sub-GeV dark matter (DM) particles in the Milky Way or macroscopic objects such as primordial black holes (PBHs) become attractive DM candidates due to null results of WIMP from direct detection experiments. We explore the possibility in which the present PBHs play as a novel source to produce light boosted DM and confine light PBHs with current and future terrestrial facilities. We study the electron elastic scattering data and obtain the current constraints from Super-Kamiokande and XENON1T on the boosted DM from PBH evaporation. The prospective bounds on the sub-GeV DM-electron scattering cross section and the fraction of DM composed of PBHs $f_{\rm PBH}$ are also imposed for future Xenon experiments.
\end{abstract}

\maketitle

\section{Introduction}
\label{sec:Intro}

Astrophysical evidence suggests that 84\% of the matter in the Universe is constituted by dark matter (DM).
However, the constitution and the properties of DM are still unknown. One longstanding candidate of thermal DM is the weakly interacting massive particle (WIMP) with weak scale interaction. Due to null conclusive evidence of WIMP search at DM direct detection (DD) experiments~\cite{Schumann:2019eaa}, however, much more efforts have been recently paid to the hypotheses beyond the WIMP in both theoretical and experimental aspects. The explorations of alternative DM candidates motivated a hypothetical light particle with sub-GeV mass~\cite{Griest:1990kh,Essig:2011nj,Essig:2017kqs,Elor:2021swj} or macroscopic objects such as primordial black holes (PBHs)~\cite{Zeldovich:1967lct,Carr:1974nx,Carr:1975qj} (see Ref.~\cite{Carr:2021bzv} for a recent review).

The DM in the Milky Way (MW) halo can be in part composed of a non-thermal light DM which gets boosted to a semi-relativistic velocity by cosmic rays~\cite{Cappiello:2018hsu,Yin:2018yjn,Bringmann:2018cvk,Ema:2018bih}, the DM semi-annihilation~\cite{DEramo:2010keq} or the annihilation of heavier DM component~\cite{Agashe:2014yua}. The boosted DM (BDM) was proposed in many well-motivated DM models and can induce high-energy recoil events by upscattering on electrons in the terrestrial experiments~\cite{Kannike:2020agf,Fornal:2020npv,Su:2020zny,Cao:2020bwd,Lei:2020mii,Bloch:2020uzh,Kannike:2020agf}.
On the other hand, it is well known that PBHs with the mass $\sim 10^{15}$ g would evaporate due to Hawking radiation~\cite{Hawking:1974rv} and cannot provide all the observed abundance of DM~\cite{Barrau:2003xp}. The emitted particles such as gamma-rays and $e^\pm$ in the evaporation process suffer from a variety of cosmological constraints~\cite{Carr:2020gox,Laha:2019ssq,Laha:2020ivk,Saha:2021pqf,Ray:2021mxu}. The neutrinos emitted from PBHs were also studied to a large extent~\cite{Dasgupta:2019cae,Wang:2020uvi,Calabrese:2021zfq,DeRomeri:2021xgy,Ghosh:2021vkt,Capanema:2021hnm,Chao:2021orr,Bernal:2022swt}.
The emitted particles can acquire energies to a few hundred MeV from the evaporation of PBHs with the mass $\sim 10^{15}$ g. The light particles can gain enough kinetic energy to travel through the Earth and reach the terrestrial detectors. Besides the Standard Model (SM) particles, there is a possibility that the present PBHs play as a novel source to produce and boost light DM~\cite{Khlopov:2004tn}. It is thus interesting to confine the PBHs through the boosted DM scattering with current and future terrestrial facilities.

In this work we explore the boosted DM from the present PBH evaporation (denoted by ``PBHBDM'' below) so as to connect the macroscopic PBHs to sub-GeV DM particle. The PBHs with mass $\sim 10^{14}-10^{16}$ g evaporate DM flux in the present Universe. The light components of DM are boosted to (semi-)relativistic velocities. Then PBHBDM travels through the Earth and scatters with the electrons in the underground detector targets of the terrestrial experiments~\footnote{The scattering of boosted DM evaporated from PBHs and nucleons was studied in Ref.~\cite{Calabrese:2021src}.}. The electrons then induce photoelectron signatures in the time-projection chambers (TPCs). Such ``S2'' signature can be sensitive to smaller DM mass region compared to the DM-nucleon scattering.

Both neutrino and DM scattering experiments can search for the PBHBDM in light of the electron elastic scattering events.
Super-Kamiokande (Super-K) with 161.9 kiloton-years exposure~\cite{Super-Kamiokande:2017dch} provide a strong constraint on sub-GeV BDM~\cite{Ema:2018bih}. The kinetic energy of the recoiled electron is above 100 MeV at Super-K.
The Xenon collaboration recently collected low-energy electron recoil data from the XENON1T experiment with an exposure of
0.65 tonne-years and reported the event distribution with a broad spectrum~\cite{XENON:2020rca}. If using the BDM event rate as well as the XENON1T data, one can construct a conservative method to give a general exclusion limit on the DM-electron scattering.
This limit then becomes a strong existing constraint on DM lighter than a few MeV. We will obtain the current constraints from Super-K and XENON1T on PBHBDM, and compute the prospects in future Xenon experiments. Moreover, these bounds can also be converted into the upper limit on the fraction of DM composed of PBHs $f_{\rm PBH}$. Compared with the current evaporation constraints from extragalactic gamma-rays, we expect the bound on $f_{\rm PBH}$ can be improved from terrestrial facilities.

This paper is organized as follows. In Sec.~\ref{sec:PBH} we evaluate the BDM spectrum from PBH evaporation. The DM-electron scattering and the scattering rate in the terrestrial facilities are then calculated in Sec.~\ref{sec:BDMelectron}. We obtain the current constraints on PBHs from Super-K and XENON1T, and give the future prospects in XENONnT in Sec.~\ref{sec:res}. Our conclusions are drawn in Sec.~\ref{sec:Con}.

\section{The DM flux from PBH evaporation}
\label{sec:PBH}

The PBHs have quantum properties and thermally radiate with a temperature~\cite{Hawking:1975vcx,Page:1976df,Page:1977um,MacGibbon:1990zk,MacGibbon:1991tj}
\begin{eqnarray}
T_{\rm PBH}= {\hbar c^3\over 8\pi G M_{\rm PBH} k_{\rm B}}\approx 10^{-7}\Big({M_{\rm PBH}\over M_\odot}\Big)^{-1} {\rm K}\;,
\end{eqnarray}
where $G$ is the Newtonian constant of gravitation, $M_{\rm PBH}$ denotes the PBH mass and $k_{\rm B}$ is the Boltzmann constant.
We use the public code BlackHawk v2.0~\cite{Arbey:2019mbc,Arbey:2021mbl}~\footnote{BlackHawk v2.0 provides the additional emission of massive DM and we directly apply it to our massive Dirac fermion DM case.} to calculate the differential number of DM per unit time emitted by PBHs~\cite{Hawking:1971ei,Page:1976df,Page:1976ki}
\begin{eqnarray}
{d^2N_\chi\over dT_\chi dt}={g_\chi\over 2\pi}{\Gamma(T_\chi,M_{\rm PBH})\over {\rm exp}((T_\chi+m_\chi)/k_{\rm B}T_{\rm PBH})+1}\;,
\end{eqnarray}
where $\Gamma$ is the function of greybody factor which encodes the probability of an
elementary spin-1/2 DM $\chi$ to escape the PBH gravitational well, $T_\chi$ is the emitted DM kinetic energy and $g_\chi=4$ denotes the degrees of freedom for the emitted DM as Dirac fermions. We next ignore the spin of PBH for simplicity~\footnote{It is customary in the literatures to only consider non-spinning PBHs (in Refs.~\cite{Wang:2020uvi,Calabrese:2021zfq,Calabrese:2021src} and etc.).
In principle, spinning PBHs evaporate faster than the non-spinning PBHs and thus can contribute to the DM content nowadays for $M_{\rm PBH}\gtrsim 7\times 10^{15}$ g~\cite{Dasgupta:2019cae}. In this work, we focus on $M_{\rm PBH}\lesssim 1\times 10^{16}$ g and only consider non-spinning PBHs.}.

\begin{figure}[htb!]
\begin{center}
\includegraphics[scale=1,width=0.9\linewidth]{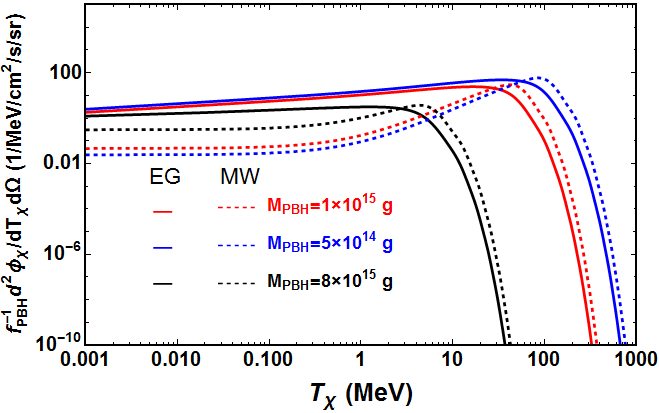}
\end{center}
\caption{The differential BDM flux from PBHs $f_{\rm PBH}^{-1}d^2\phi_\chi/dT_\chi d\Omega$ as a function of kinetic energy $T_\chi$ for three benchmark values of $M_{\rm PBH}=5\times 10^{14}$ g (blue), $1\times 10^{15}$ g (red), and $8\times 10^{15}$ g (black). The EG and MW contributions are denoted by solid and dashed lines, respectively. We fix $m_\chi=1$ MeV.
}
\label{fig:BDMflux}
\end{figure}

For the DM flux from PBH evaporation, we consider the contributions of both the PBHs in galactic halo and
the extragalactic PBHs~\cite{Wang:2020uvi,Calabrese:2021zfq,DeRomeri:2021xgy}
\begin{eqnarray}
{d^2\phi_\chi\over dT_\chi d\Omega}={d^2\phi_\chi^{\rm MW}\over dT_\chi d\Omega}+{d^2\phi_\chi^{\rm EG}\over dT_\chi d\Omega}\;,
\label{eq:PBHnu}
\end{eqnarray}
where $\phi_\chi^{\rm MW}$ and $\phi_\chi^{\rm EG}$ correspond to the DM flux from PBHs in Milky Way (MW) and extragalactic (EG) PBHs, respectively, and $\Omega$ is the solid angle.
The differential galactic DM flux is given by
\begin{eqnarray}
{d^2\phi_\chi^{\rm MW}\over dT_\chi d\Omega}={f_{\rm PBH}\over 4\pi M_{\rm PBH}} {d^2N_\chi\over dT_\chi dt} \int {d\Omega_s\over 4\pi}\int dl \rho_{\rm MW}[r(l,\psi)]\;,
\label{eq:MWnuflux}
\end{eqnarray}
where $f_{\rm PBH}$ denotes the fraction of DM composed of PBHs, $\rho_{\rm MW}[r(l,\psi)]$ is the DM density of the MW halo, $r(l,\psi)=\sqrt{r_\odot^2+l^2-2lr_\odot \cos\psi}$ is the galactocentric distance with $r_\odot$ the solar distance from the galactic center, $l$ the line-of-sight distance to the PBH and $\psi$ the angle between these two directions. The maximal limit of distance $l$ is taken to be $(r_h^2-r_\odot^2 \sin^2\psi)^{1/2}+r_\odot \cos\psi$ with the halo radius being $r_h=200$ kpc~\cite{Wang:2020uvi}. The angular integration is defined as $\int d\Omega_s=\int_0^{2\pi}d\phi \int_0^\pi d\psi \sin\psi$ with the azimuthal angle $\phi$. For illustration, we employ the generalized Navarro-Frenk-White (NFW) DM profile~\cite{Navarro:1996gj}
\begin{eqnarray}
\rho_{\rm MW}(r)=\rho_\odot \Big({r\over r_\odot}\Big)^{-\gamma} \Big({1+r_\odot/r_s\over 1+r/r_s}\Big)^{3-\gamma}\;,
\end{eqnarray}
where $\rho_\odot=0.4~{\rm GeV}/{\rm cm}^3$ is the local DM density, $r_\odot=8.5$ kpc, $r_s=20$ kpc is the radius of the galactic diffusion disk, and the inner slope of the NFW halo profile is fixed as $\gamma=1$. For the extragalactic contribution, the corresponding differential BDM flux over the full sky is
\begin{eqnarray}
{d^2\phi_\chi^{\rm EG}\over dT_\chi d\Omega}={f_{\rm PBH}\rho_{\rm DM}\over 4\pi M_{\rm PBH}}\int^{t_{\rm max}}_{t_{\rm min}} dt [1+z(t)]{d^2N_\chi\over dT_\chi dt}\Big|_{E^s_\chi}\;,
\label{eq:EGnuflux}
\end{eqnarray}
where the average DM density of the Universe at the
present epoch determined by Planck~\cite{Planck:2018vyg} is $\rho_{\rm DM}=2.35\times 10^{-30}~{\rm g}/{\rm cm}^3$, $E^s_\chi=\sqrt{(E_\chi^2-m_\chi^2)(1+z(t))^2+m_\chi^2}$ denotes the energy at the source which is related to the energy $E_\chi$ in the observer's frame by the redshift $z(t)$. The DM emitted from the PBHs in the very early Universe is sufficiently redshifted.
As a result, their energies are very low today and cannot be detected. Thus, for the integral limits, we fix $t_{\rm min}=10^{11}$ s being close to the era of matter-radiation equality~\cite{Wang:2020uvi} and $t_{\rm max}=\tau_0$ with $\tau_0$ as the age of Universe. Actually, it turns out that changing the lower limit has no impact on the results~\cite{DeRomeri:2021xgy}. In Fig.~\ref{fig:BDMflux} we show the differential BDM flux from PBHs $f_{\rm PBH}^{-1}d^2\phi_\chi/dT_\chi d\Omega$ for three benchmark values of $M_{\rm PBH}$. As seen in Eqs.~(\ref{eq:MWnuflux}) and (\ref{eq:EGnuflux}), smaller PBHs exhibit harder spectra of the evaporated DM.
The PBHBDMs have maximal kinetic energies of order $T_\chi\sim \mathcal{O}(100)$ MeV for $M_{\rm PBH}\lesssim 10^{15}$ g.
The spectrum generally peaks at lower energies for each PBH mass. The energy where the spectrum peaks at is higher for less massive PBH.

\section{PBHBDM-electron scattering in the terrestrial facilities}
\label{sec:BDMelectron}

The produced BDM $\chi$ can travel a distance underground and scatter with the electrons in the detector of the terrestrial facilities. After traveling a distance of $z$, the actual DM flux reaching the detector is given by~\cite{Calabrese:2021src}
\begin{eqnarray}
	{d^2\phi_\chi^d\over dT_\chi^d d\Omega}\approx {4m_\chi^2 e^\tau\over (2m_\chi+T_\chi^d-T_\chi^d e^\tau)^2}{d^2\phi_\chi\over dT_\chi d\Omega}\Big|_{T^0_\chi}\;,
\label{eq:DMflux}
\end{eqnarray}
where $T_\chi^0$ and $T_\chi^d$ are defined as the DM kinetic energies at Earth's surface
and at the detector, respectively. They are related as
\begin{eqnarray}
	T^0_\chi(T^d_\chi)={2m_\chi T^d_\chi e^\tau\over 2m_\chi+T^d_\chi-T^d_\chi e^\tau}\;,
\end{eqnarray}
where $\tau=z/\ell_{\rm int}^\oplus$\,,
with
\begin{eqnarray}
z=-(R-d)\cos\theta_z+\sqrt{R^2-(R-d)^2\sin^2\theta_z}\,,
\end{eqnarray}
and
\begin{eqnarray}
	\ell_{\rm int}^\oplus=\Big[n^\oplus_e\sigma_{\chi e}{2m_e m_\chi\over (m_e+m_\chi)^2}\Big]^{-1}\;.
\end{eqnarray}
Here $R$ is the radius of the Earth, $\theta_z$ is the zenith angle of the detector, $d$ is the depth of the detector from the Earth surface at the zero zenith angle, $\sigma_{\chi e}$ is the total cross section of BDM-electron scattering, and $n^\oplus_e$ is the electron number density of the Earth. For simplicity, we take the average value over the Earth, i.e., $n^\oplus_e=8\times10^{23}~\text{cm}^{-3}$~\cite{Ema:2018bih}.
Note that the BDM can also be stopped by the scattering in the atmosphere~\cite{Emken:2018run,Bringmann:2018cvk,Ema:2018bih,Cappiello:2019qsw} and there would be an additional $z/\ell_{\rm int}^{\rm atm}$ due to atmosphere attenuation in quantity $\tau$. To estimate the electron number density in the atmosphere, we consider the density profile of the US Standard Atmosphere model with an altitude of $d^{\rm atm}=86$ km~\cite{Emken:2018run}. The atmosphere was modeled as 4 layers of falling mass density as a function of the altitude. We take a constant density $\rho=1.1~{\rm kg}~{\rm m}^{-3}$ in the first layer and the most abundant
elements in the atmosphere (nitrogen and oxygen) to obtain the most optimistic number density of electron as $n_e^{\rm atm}=3.2\times 10^{20}~{\rm cm}^{-3}$. For zero zenith angle, we find that the value of $d^{\rm atm}n^{\rm atm}_e$ in $\tau$ is 30 and 40 times smaller than $dn^\oplus_e$ of Super-K and XENON1T detectors, respectively.
Thus, the atmosphere attenuation is negligible and we neglect it in the following analysis.

After combining the above DM flux, the number of scattered DM events per unit time per solid angle per
recoil energy $T_e$ of the electron, is then given by
\begin{eqnarray}
	{d^3 N_\chi\over dt d\Omega dT_e}=N_e\int dT_\chi^d  \sigma_{\chi e}D^e_\chi(T_e,T_\chi^d) {d^2\phi_\chi^d\over dT_\chi^d d\Omega}\;,
	\label{eq:dN}
\end{eqnarray}
where $N_e$ is the total number of free electrons in targets for a detector.
We define a transfer function which encodes the electron energy spectrum induced by the scattering with BDM $\chi$ as
\begin{eqnarray}
	D^e_\chi(T_e,T_\chi)={1\over T_e^{\rm max}(T_\chi)}\Theta(T_e^{\rm max}(T_\chi)-T_e)\;,
\end{eqnarray}
where the final kinetic energy of electrons is
\begin{eqnarray}
	T_e=T_e^{\rm max}{1-\cos\theta\over 2}\;,
\end{eqnarray}
with
\begin{eqnarray}
	T_e^{\rm max}={2m_e(T_\chi^2+2m_\chi T_\chi)\over (m_\chi+m_e)^2+2m_e T_\chi}\;.
\end{eqnarray}
Here $\theta$ is the scattering anle in the center of mass (CM) frame, and we assume the scattering is isotropic in the CM frame.

\begin{figure}[htb!]
	\begin{center}
		\includegraphics[scale=1,width=0.8\linewidth]{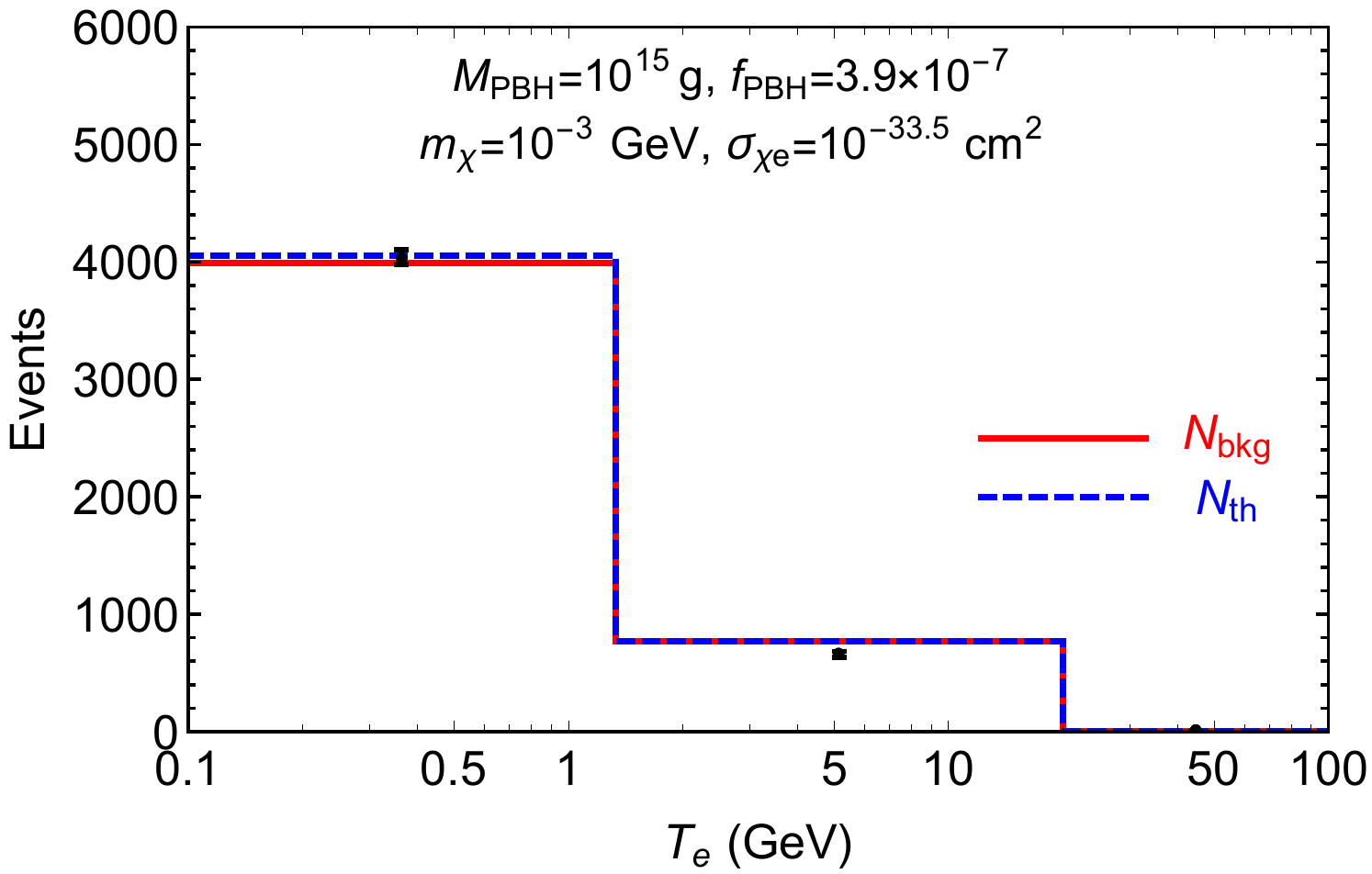}
		\includegraphics[scale=1,width=0.8\linewidth]{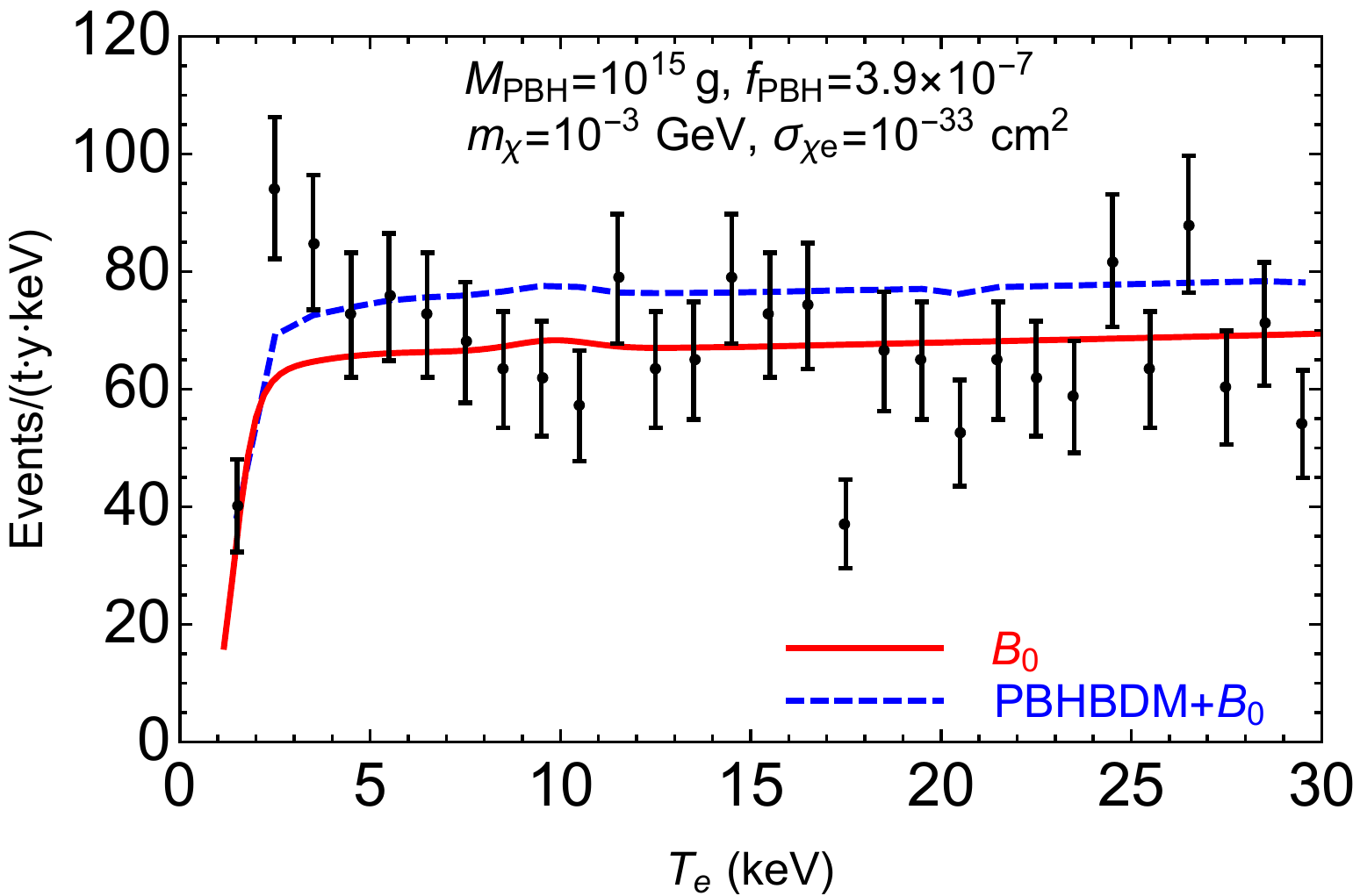}
	\end{center}
	\caption{Top: the predicted event spectrum (blue dashed) and the estimated backgrounds (red solid) at Super-K. For the predicted events, we set $\alpha_1=-0.35$, $\alpha_2=0$ and $\alpha_3=0$.  Bottom: the electron recoil spectrum in XENON1T induced by PBHBDM with (blue dashed) the background model spectrum $B_0$ (red solid).
	}
	\label{fig:events}
\end{figure}

\subsection{Super-Kamiokande}
\label{sec:superk}
\begin{table}
	\centering
	\begin{tabular}{| c |  c |  c |  c | }
		\hline
		Energy bin  & Data& Background  & Uncertainty \\
		\hline
		$0.1\text{ GeV} < T_e<1.33\text{ GeV} $  & 4042& 3992.9 &19\%\\
		\hline
		$1.33\text{ GeV} < T_e<20\text{ GeV} $    & 658 & 772.6&19\% \\
		\hline
		$T_e>20\text{ GeV}$    &  3& 7.4& 23\%  \\
		\hline
	\end{tabular}
	\caption{The number of measured events, estimated backgrounds and systematic uncertainties for each electron recoil energy bin at Super-K~\cite{Super-Kamiokande:2017dch}. }
	\label{Table: SK}
\end{table}

To set bounds on the PBHBDM parameter space from Super-K, we use the ``electron elastic scatter-like'' events from Ref.~\cite{Super-Kamiokande:2017dch}, and define the $\chi^2$ as
\begin{eqnarray}
	\chi^2=\sum_{i=1}^3 \min_{\alpha_i} \left[2\left(N_\text{th}^i-N_\text{obs}^i+N_\text{obs}^i\ln\frac{N_\text{obs}^i}{N_\text{th}^i}\right)+\left(\frac{\alpha_i}{\sigma_i}\right)^2\right]\,,\nonumber \\
\end{eqnarray}
where the predicted number of events per bin $N_\text{th}^i=N_\text{bkg}^i(1+\alpha_i)+N_\text{sig}^i$. Here $N_\text{bkg}^i$ ($N_\text{obs}^i$) [$N_\text{sig}^i$] is the number of background (observed) [signal] events in each bin, $\alpha_i$ is the background normalization factor that is minimized over, and $\sigma_i$ is the uncertainty for each bin. The main background at Super-K is due to atmospheric neutrinos~\cite{Super-Kamiokande:2017dch}. Follow Ref.~\cite{Super-Kamiokande:2017dch}, we take a conservative value of $19\%$ for the uncertainties in the two lower energy bins, and 23\% for the uncertainty in the highest energy bin. The number of measured event, estimated background and the uncertainty for each energy bin are listed in Table~\ref{Table: SK}.
For the signal events, we take into account the Earth attenuation effect in the calculation of the predicted number of events from PBHBDM.
Since the distance between the detector and the surface of the Earth depends on the direction of observation, we also integrate over the solid angle for the Super-K detector. The predicted number of events is obtained by integrating Eq.~(\ref{eq:dN}) over the solid angle with the data-taking time of 2628.1 days and  $N_e^\text{SK}=7.5\times10^{33}$, and we also apply the signal efficiency extracted from Fig.~1 in Ref.~\cite{Super-Kamiokande:2017dch}.
For illustration, a predicted event spectrum for $M_{\rm PBH}=10^{15}$ g, $f_{\rm PBH}=3.9\times10^{-7}$, $\sigma_{\chi e}=10^{-33.5}~{\rm cm}^2$, $m_\chi=10^{-3}$ GeV with $\alpha_1=-0.35$, $\alpha_2=0$ and $\alpha_3=0$ at Super-K is shown in the upper panel of Fig.~\ref{fig:events}. The number of PBHBDM events drops very fast as the electron recoil energy increases, and there are almost no PBHBDM events at the higher energy bins.
The background-only hypothesis yields a $\chi^2_{0}=2.7$, and the 2$\sigma$ exclusion bound is set with $\Delta \chi^2=\chi^2-\chi^2_{0}=4.0$  for one degree of freedom.

\subsection{XENON1T and prospects}
\label{sec:xenon1t}

We also use the XENON1T electron recoil data measured in Ref.~\cite{XENON:2020rca} to set bounds on PBHBDM. After taking account of the Earth attenuation effect, the differential number of events from PBHBDM is given by
\begin{eqnarray}
{d N_\chi\over dT_e}= Z_\text{eff}^\text{Xe}(T_e){M_\text{det} t\over m_\text{Xe}} \int  d\Omega dT_\chi^d  \sigma_{\chi e} D^e_\chi(T_e,T_\chi^d) {d^2\phi^d_\chi\over dT^d_\chi d\Omega}   \;,\nonumber \\
\end{eqnarray}
where $M_\text{det}$ is the fiducial mass of the detector, $t$ is the exposure time, $m_\text{Xe}$ is the atomic mass of Xenon, and $Z_\text{eff}^\text{Xe}$ is the effective electron charge of Xenon that depends on the electron recoil energy $T_e$. Here we take the values of $Z_\text{eff}^\text{Xe} (T_e)$ from the Appendix A of Ref.~\cite{AtzoriCorona:2022jeb}, which are obtained by using the edge energies extracted from the photoabsorption data~\cite{Chen:2016eab,Henke:1993eda}.
To compare with the measured data in Fig.~4 of Ref.~\cite{XENON:2020rca}, we take $M_\text{det} t = 1$ ton$\cdot$year, and convolve the differential number of events with a Gaussian detector response function for the electron recoil spectrum
\begin{eqnarray}
	\sigma(E)=a\sqrt{E}+bE\;,
\end{eqnarray}
where $a=0.31~\sqrt{\rm keV}$ and $b=0.0037$~\cite{XENON:2020rca}. We further apply the detector efficiency and add the background $B_0$ from Ref.~\cite{XENON:2020rca} to the PBHBDM contribution. The background components at XENON1T come from solar neutrinos, $\gamma$ emissions from radioimpurities in detector materials, $\beta$ decay of $^{214}$Pb, $^{85}$Kr, and $^{133}$Xe, internal conversion of $^{131\rm m}$Xe and $^{83\rm m}$Kr, and electron capture of $^{125}$I and $^{124}$Xe~\cite{XENON:2020rca}. To place the bound on PBHDM parameter space, we define a simple test statistic as follows
\begin{eqnarray}
	\chi^2=\sum_i \frac{\left[\frac{dN_{\chi+B_0,i}}{dT_\text{rec}}-\frac{dN_{obs, i}}{dT_\text{rec}}\right]^2}{\sigma_i^2}\,,
\end{eqnarray}
where $T_\text{rec}$ is the reconstructed energy after the smearing, $\frac{dN_{obs, i}}{dT_\text{rec}}$ and $\sigma_i$ are the observed event rate and uncertainty in the $i^\text{th}$ bin extracted from Fig.~4 in Ref.~\cite{XENON:2020rca}, respectively. We find the background-only hypothesis yields a $\chi^2_{B_0}=46.4$. The 2$\sigma$ bound is estimated with $\Delta \chi^2=\chi^2-\chi^2_{B_0}=4.0$  assuming the test
statistic follows a chi-squared distribution with one degree of freedom.

For the projected sensitivity at the future XENONnT experiment~\cite{XENON:2020kmp}, we consider a detector with 20 ton$\cdot$year exposure. We use the same binning as the measured spectrum at XENON1T and perform a $\chi^2$ analysis with
\begin{eqnarray}
	\chi^2=\sum_i \frac{N_{\chi,i}^2}{N_{B,i}}\;,
\end{eqnarray}
where $N_{\chi,i}$ ($N_{B,i}$) is the expected number of signal (background) events
in the $i^\text{th}$ bin at XENONnT. The detection efficiency and the energy resolution are assumed to be the same as those in the current XENON1T detector, and the number of background events per ton$\cdot$year are reduced by a factor of 6 at XENONnT~\cite{XENON:2020rca, XENON:2020kmp}.

\section{Results}
\label{sec:res}
\begin{figure}[htb!]
\begin{center}
\includegraphics[scale=1,width=0.95\linewidth]{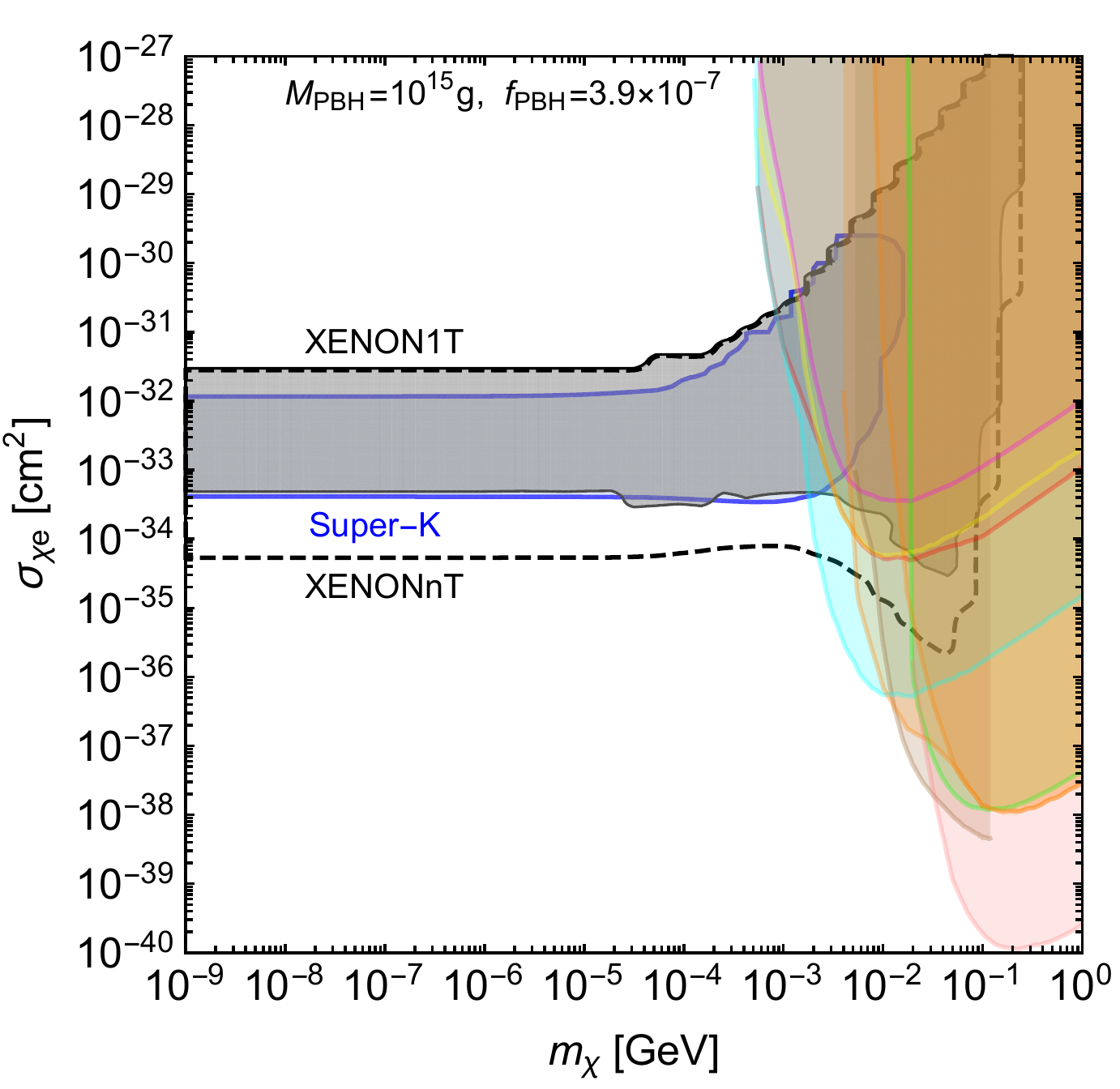}
\end{center}
\caption{$2\sigma$ bounds from Super-K (shaded blue) and XENON1T (shaded gray) and projected sensitivity at XENONnT (black dashed line) on $\sigma_{\chi e}$ as a function of $m_\chi$ for $M_{\rm PBH}=10^{15}$ g and $f_{\rm PBH}=3.9\times10^{-7}$. The constraints from DM direct detection experiments are also shown, including SENSEI~\cite{SENSEI:2020dpa} (cyan), DAMIC~\cite{DAMIC:2019dcn} (red), EDELWEISS~\cite{EDELWEISS:2020fxc} (yellow), CDMS~\cite{SuperCDMS:2018mne} (magenta),  DarkSide50~\cite{DarkSide:2018ppu} (green), XENON10/100~\cite{Essig:2017kqs} (orange), XENON1T~\cite{XENON:2019gfn} (pink) and PandaX-II~\cite{PandaX-II:2021nsg} (brown).
}
\label{fig:limit}
\end{figure}

We first show the bounds on the $\sigma_{\chi e}$ versus $m_\chi$  parameter space from Super-K and XENON1T in Fig.~\ref{fig:limit} assuming $M_{\rm PBH}=1\times 10^{15}$ g and $f_{\rm PBH}=3.9\times 10^{-7}$ allowed by the current constraints from extragalactic gamma-rays~\cite{Carr:2020gox}.
From Fig.~\ref{fig:limit}, one can see that
the DM-electron cross section between $5\times10^{-34}$ ($4\times10^{-34}$) and $3\times10^{-32}~{\rm cm}^2$ ($1\times10^{-32}~{\rm cm}^2$) is excluded by XENON1T (Super-K) for light $m_\chi\lesssim 1$ MeV. Above this region, the cross section $\sigma_{\chi e}$ is too large to make the BDM reach the detector.
We also estimate the projected sensitivity on $\sigma_{\chi e}$ as a function of $m_\chi$ at future XENONnT experiment with 20 ton$\cdot$year exposure~\cite{XENON:2020kmp}.
As we see from Fig.~\ref{fig:limit}, the future XENONnT experiment can improve the current lower bounds by about a factor of 3.
The constraints on the $\sigma_{\chi e}-m_\chi$ parameter space from various DM direct detection experiments are also added for comparison, and we see that the regions to the right of the curves have been already excluded by existing DM direct detection experiments.

\begin{figure}[htb!]
\begin{center}
\includegraphics[scale=1,width=1\linewidth]{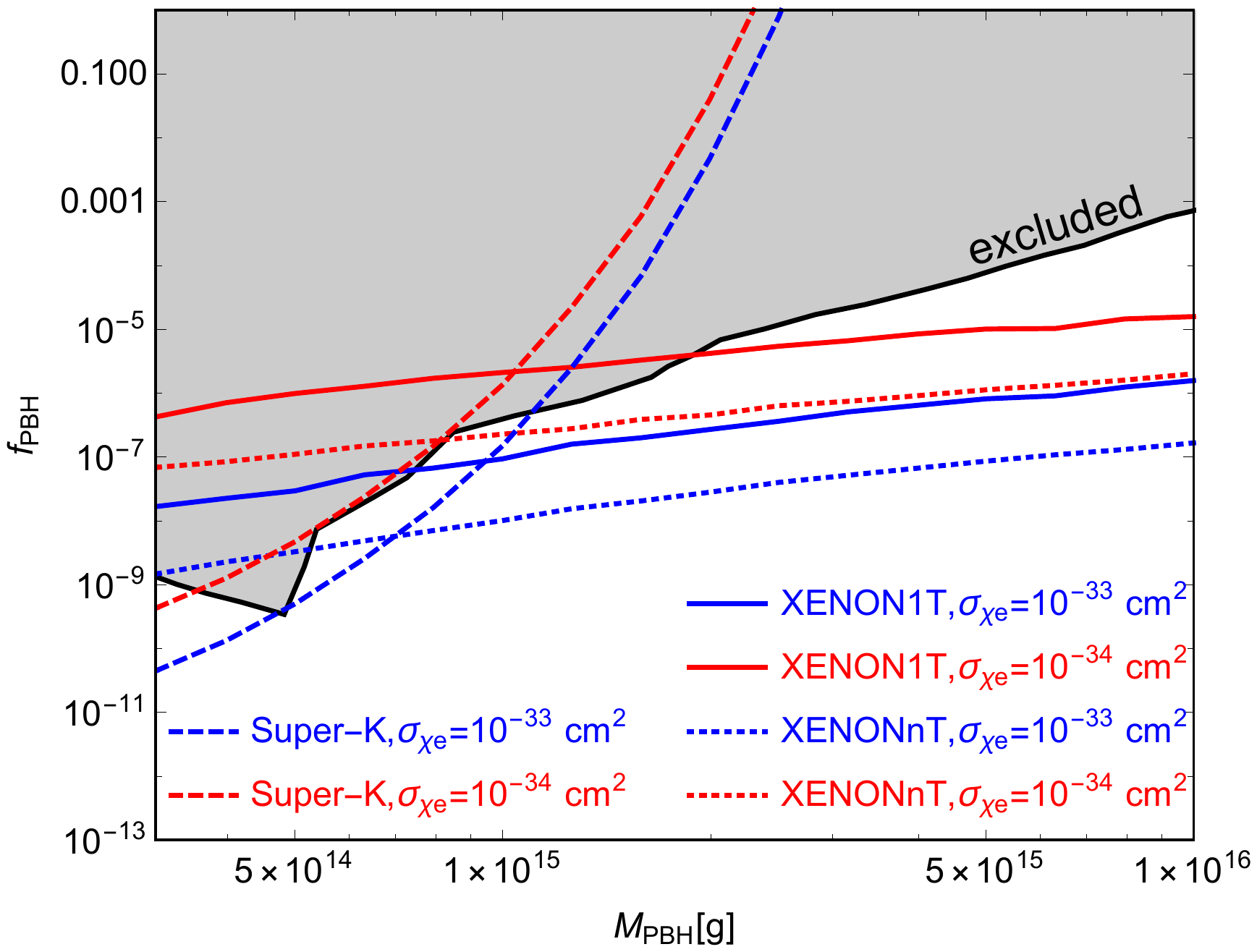}
\end{center}
\caption{$2\sigma$ bounds on $f_\text{PBH}$ as a function of $M_\text{PBH}$ for different DM-electron cross sections $\sigma_{\chi e}=10^{-33}~{\rm cm}^2$ (blue) and $\sigma_{\chi e}=10^{-34}~{\rm cm}^2$ (red), and $m_\chi=10^{-3}$ GeV. The bounds and prospect are from Super-K (dashed), XENON1T (solid) and XENONnT (dotted). The current evaporation constraints from extragalactic gamma-rays~\cite{Carr:2020gox} are also shown for comparsion.
}
\label{fig:PBH}
\end{figure}

These bounds can also be translated into the upper limit on $f_{\rm PBH}$ as a function of $M_{\rm PBH}$, which is shown in Fig.~\ref{fig:PBH} for different DM-electron cross sections and $m_\chi=10^{-3}$ GeV. As we see from Fig.~\ref{fig:PBH}, the bounds from Super-K and XENON experiments are complementary to each other. For illustration, assuming $\sigma_{\chi e}=10^{-33}~{\rm cm}^2$, Super-K sets a stronger bound than XENON1T (XENONnT) for $M_\text{PBH}\lesssim 10^{15}$ g ($M_\text{PBH}\lesssim 7\times 10^{14}$ g). Compared with the current evaporation constraint from extragalactic gamma-rays, an overall improvement of $f_{\rm PBH}$ bound can be achievable for Super-K and XENON1T combined.

\section{Conclusions}
\label{sec:Con}

The macroscopic PBH and the light sub-GeV DM particle are two interesting DM candidates beyond the WIMP. The boosted DM from PBH evaporation can be searched at both neutrino and DM direct detection experiments.
Using the current data from Super-K and XENON1T, we place bounds on the DM-electron cross section for sub-GeV DM. We find that for light $m_\chi\lesssim 1$ MeV, assuming $f_{\rm PBH} = 3.9 \times 10^{-7}$ and $M_{\rm PBH} = 10^{15}$ g, the DM-electron cross section between $5\times10^{-34}$ ($4\times10^{-34}$) and $3\times10^{-32}~{\rm cm}^2$ ($1\times10^{-32}~{\rm cm}^2$) can be excluded by current XENON1T (Super-K) data.
Future XENONnT experiment can improve the current lower bounds by about a factor of 3. We also find the bounds from Super-K and XENON on the $f_\text{PBH}$ versus $M_\text{PBH}$ parameter space are complementary to each other.
Compared with the current evaporation constraints from extragalactic gamma-rays, the bound on $f_{\rm PBH}$ from Super-K and XENON1T/XENONnT can be substantially improved with a certain range of DM-electron cross section for light DM particles.
Note that these bounds on $f_{\rm PBH}$ and $\sigma_{\chi e}$
are obtained simultaneously and are fully correlated.

\acknowledgments
T.L. would like to thank Roberta Calabrese, Wei Chao and Xin-He Meng for useful discussions.
T.L. is supported by the National Natural Science Foundation of China (Grant No. 11975129, 12035008) and ``the Fundamental Research Funds for the Central Universities'', Nankai University (Grant No. 63196013). J.L. is supported by the National Natural Science Foundation of China (Grant No. 11905299), Guangdong Basic and Applied Basic Research Foundation (Grant No. 2020A1515011479), the Fundamental Research Funds for the Central Universities, and the Sun Yat-Sen University Science Foundation.

\vspace{0.1 in}
{\bf Note added.} After the completion of this work, a separate work on constraints on light dark matter produced in primordial black hole evaporation from electron-target experiments has appeared~\cite{Calabrese:2022rfa}. Their bounds for a fixed $\sigma_{\chi e}$ are similar to our Fig.~\ref{fig:PBH}.
\vskip1cm

\appendix

\bibliography{refs}

\end{document}